\documentstyle{article}
\font \msb=msbm10 scaled \magstep1

\newcommand{\bC}{\mbox{\msb C} }

\def\a{{\alpha}}
\def\b{{\beta}}
\def\g{{\gamma}}
\def\d{{\delta}}
\def\e{{\varepsilon}}
\def\s{{\sigma}}
\def\A{{\cal A}}
\def\B{{\cal B}}
\def\C{{\cal C}}
\def\D{{\cal D}}
\def\F{{\cal F}}
\def\L{{\cal L}}
\def\U{{\cal U}}
\def\T{{\cal T}}
\def\H{{\cal H}}
\def\f{\mbox{\bf f}}
\def\e{\mbox{\bf e}}
\def\h{\mbox{\bf h}}
\def\te#1{{\widetilde{#1}}}

\def\on#1#2{\mathop{\vbox{\ialign{##\crcr\noalign{\kern2pt}
$\scriptstyle{#2}$\crcr\noalign{\kern2pt\nointerlineskip}
\kern-2pt$\hfil\displaystyle{#1}\hfil$\crcr}}}\limits}
\def\tr{{\rm tr}\,}
\def\nn{ \nonumber }
\def\bq{ \begin{equation} }
\def\eq{ \end{equation} }
\def\ben{ \begin{eqnarray} }
\def\en{ \end{eqnarray} }
\def\ll{ \label }
\def\dfrac#1#2{{\displaystyle{#1\over#2}}}
\def\frac#1#2{{{#1\over#2}}}

\newtheorem{pr}{Proposition}

\begin{document}
\begin{titlepage}
\title{
\vskip2cm
{\flushright
\small\bf Preprint solv-int/9801012
\flushleft}
\vskip2cm
Dynamical boundary conditions for integrable lattices.}
\author{
 A.V. Tsiganov\\
{\small\it
 Department of Mathematical and Computational Physics,
 Institute of Physics,}\\
{\small\it
St.Petersburg University,
198 904,  St.Petersburg,  Russia}
}
\date{}
\maketitle
\thispagestyle{plain}

\begin{abstract}
Some special solutions to the reflection equation are
considered.  These boundary matrices are defined on the common
quantum space with the other operators in the chain.  The
relations with the Drinfeld twist are discussed.
\end{abstract}

\end{titlepage}
\vfill
\newpage

\section{On a covariance property of the reflection equation}
\setcounter{equation}{0}
Let the operator-valued function $R(u):\bC\to{\rm Aut}(V\otimes
V)$ be a solution to the quantum Yang-Baxter equation
\cite{fld84} in a finite-dimensional linear space $V$.  Let us
connect with a given matrix $R(u)$ two associative algebras
$\T_R$ and $\U_R$ generated by noncommutative entries of the
square matrices $T(u)$ and $K(u)$ satisfying the fundamental
commutator relation \cite{fld84,ks82}
\bq
R_{12}(u-v)\,{\on{T}1}(u)\,{\on{T}2}(v)=
{\on{T}2}(v)\,{\on{T}1}(u)\,R_{12}(u-v)\,, \ll{fcr}
\eq
or the reflection equation \cite{cher84,skl88}
\bq
R_{12}(u-v)\,{\on{K}{\,1}}(u)\,R_{21}(u+v)\,{\on{K}{\,2}}(v)=
{\on{K}{\,2}}(v)\,R_{12}(u+v)\,{\on{K}{\,1}}(u)\,R_{21}(u-v)\,.
\ll{re}
\eq
Here  $\on{X}{1}\equiv X \otimes id_{V_2}$, $\on{X}{2}\equiv
id_{V_1} \otimes X$ for any matrix $X\in End(V) $. As usual,
$R_{ij}(u)\in{\rm End}(V_i\otimes V_j)\,,~V_j\equiv V$ and
$R_{ji}(u)=PR_{ij}(u)P$, with $P$ as the permutation operator in
the tensor product of the two spaces $V_i\otimes V_j$
\cite{fld84,ks82}.

In this paper we discuss solutions to the reflection equation
(\ref{re}), i.e. consider various representations of
the algebra $\U_R$ \cite{skl88}.  There is plenty of
known representations \cite{cher84,skl88,kuzts89a,dv94}.
So, the covariance property of the reflection equation
(\ref{re}) \cite{skl88,kusas93} may be used to construct new
solutions starting from known ones.
Namely, it was pointed out in \cite{skl88} that
\begin{pr}
Let the matrices $T(u)$ and $K(u)$ satisfy the
relations (\ref{fcr}) and (\ref{re})
with the same $R$-matrix $R(u)$, then
the Sklyanin monodromy matrix
\bq
K'(u)=T(u)K(u)T^{-1}(-u)\,,\ll{t-}
\eq
solves the reflection equation (\ref{re}) if
\bq \on{T}{1}(u)\on{K}{\,2}(v)=\on{K}{\,2}(v)\on{T}{1}(u)\,.\ll{res1}\eq
\end{pr}
The proof follows easily by the substitution  $K'(u)$ into
the reflection equation (\ref{re}) and by using few times different forms
of the fundamental relation (\ref{fcr}) e.g.
\bq
{\on{T^{-1}}2}\,(-v)R_{12}(u+v)\,{\on{T}1}(u)=
{\on{T}1}(u)\,R_{12}(u+v)\,{\on{T^{-1}}2}(-v)\,.\ll{fcr2}
\eq
The main condition (\ref{res1}) holds if $T(u)$ and $K(u)$ are
some representations of the algebras $\T_R$ and $\U_R$ in the
different spaces $\H_1\otimes V$ and $\H_2\otimes V$
respectively, whence entries of $T(u)$ and $K(u)$ are mutually
commute. Here $\H_1$ and $\H_2$ are the distinct quantum
spaces and $V$ means  a common auxiliary space
\cite{fld84,ks82}.

In this paper the Sklyanin monodromy matrix $K'$ (\ref{t-}) is
constructed from the known representations $T(u)$ and $K(u)$ of
the algebras $\T_R$ and $\U_R$ defined on the common quantum
space $\H$, i.e. using matrices with the non-ultralocal
commutation relations \cite{afs92}.  In that case initial
matrix $K(u)$ will be termed the {\it dynamical} boundary matrix.

Let us preserve construction of the Sklyanin representation $K'$
(\ref{t-}) and refuse on the covariance property of the initial
reflection equation (\ref{re}).
\begin{pr}
Let $T(u)$ satisfies the fundamental commutator relation
(\ref{fcr}). If it is intertwined with dynamical boundary
matrix $K(u)$ by initial matrix $R(u)$ and by some new matrix
$S(u,v)$
\bq
\on{K}{\,1}(u)\on{T}{2}(v)~R_{12}(u+v)~\,{\on{T^{-1}}1}(-u)\on{K}{\,2}(v)=\
\on{T}{2}(v)\on{K}{\,1}(u)~S_{21}(u,v)~\on{K}{\,2}(v)\,{\on{T^{-1}}1}(-u)\,,
\ll{res4}
\eq
then the Sklyanin monodromy matrix
\[
K'(u)=T(u)K(u)T^{-1}(-u)\,,
\]
solves initial reflection equation (\ref{re}) provided that
the dynamical boundary matrix $K(u)$ satisfies the
generalized reflection equation
\bq
R_{12}(u-v)\,{\on{K}{\,1}}(u)~S_{21}(u,v)~{\on{K}{\,2}}(v)=
{\on{K}{\,2}}(v)~S_{12}(u,v)~{\on{K}{\,1}}(u)\,R_{21}(u-v)\,.
\ll{re2}
\eq
\end{pr}
While $S(u,v)$ is an arbitrary matrix and the proof consists in
direct verification of the reflection equation (\ref{re}) by
using relations (\ref{fcr},\ref{res4},\ref{re2}).

For a given matrix $T(u)$ we can try to find dynamical
boundary matrix $K(u)$ together with the matrix $S(u,v)$ from
the equation (\ref{res4}), then we have to check the
generalized reflection equation (\ref{re2}) intertwining both
these matrices with the $R$-matrix.

Particular solutions to the equation (\ref{res4})
may be obtained from the following system of equations
\ben
\on{K}{1}(u)\on{T}{2}(v)&=&\on{T}{2}(v)\on{K}{1}(u)\,F(u,v)\,,\nn\\
\ll{res3}\\
{\on{T^{-1}}1}(-u)\on{K}{2}(v)&
=&G(u,v)\,\on{K}{2}(v)\,{\on{T^{-1}}1}(-u)\,,\nn
\en
with the two unknown matrices $F(u,v)$ and $G(u,v)$.
In that case dynamical boundary matrix $K(u)$ has to be
solution to generalized reflection equation (\ref{re2})
with the matrix
\bq
S(u,v)=F(u,v)\,R(u+v)\,G(u,v)\,.\ll{stw}
\eq
Equations (\ref{res3}) have the form of the exchange algebras
\cite{afld91}.  Obviously, other form of these algebras are
\ben
F(u,v)&=&\on{K^{-1}}{\,1}(u)\on{T^{-1}}{2}(v)\,
\on{K}{1}(u)\on{T}{2}(v)\,\nn\\
\nn\\
G(u,v)&=&{\on{T^{-1}}1}(-u)\on{K}{2}(v)\,
{\on{T}1}(-u)\on{K^{-1}}{\,2}(v)\,.\nn
\en
Let the dynamical boundary matrix $K(u)$ has
the following property
\bq K(u)K(-u)=\phi(u) I \,,\ll{trpr}
\eq
where $\phi(u)$ means some scalar function. In that case
matrix $S(u,v)$ (\ref{stw}) is the Drinfeld twist \cite{d90}
of the matrix $R(u+v)$
\bq
S(u,v)=F(u,v)R(u+v)F_{21}^{-1}(-v,-u)\,,\qquad
F_{21}(u,v)=PF(u,v)P\,,\ll{dtw}
\eq
if the universal twist element  $F(u,v)$ has an
appropriate properties \cite{d90}. Note, a twist transformation
(\ref{dtw}) of the $R$-matrix related to the braid group
${\check R}= PR$
\[ {\check S}(u,v)=PS(u,v)=F_{21}(u,v){\check
R}F_{21}^{-1}(-v,-u)\,, \]
looks just like a Sklyanin formulae (\ref{t-}), in contrast with the
usual similarity transformation in the quantum group theory.

Recall, for the integrable lattice models matrix $T(u)$ is
constructed as an ordered product
\bq
T(u)=L_n(u)L_{n-1}(u)\cdots L_1(u)\,,\ll{tl}
\eq
of $n$ independent $L$-operators having some simple dependence
on spectral parameter $u$.  Matrices $L_k(u)$ in the chain
(\ref{tl}) are operators which act on the different local
spaces $\H_k\otimes V$, now we use the notation $\H_k$ for a
local quantum space assigned to the site $k$ in the lattice and
$V\simeq\bC^n$ is a common auxiliary space \cite{fld84,ks82}.

Let $K(u)$ be representation of the reflection equation algebra
$\U_R$ in the space $\H_-\otimes V$.  Then, the Sklyanin
monodromy matrix (\ref{t-}) \cite{skl88,kuzts89a,dv94}
\bq
K_-(u)=L_n(u)L_{n-1}(u)\cdots L_1(u)K(u)
L_1^{-1}(-u)L_2^{-1}(-u)\cdots L_n^{-1}(-u) \ll{sklm}
\eq
describes a lattice model with boundary conditions
corresponding to the matrix $K(u)$.  As usual,
transfer matrix is given by
\bq \tau(u)=\tr K_+(u)T(u)K(u)T^{-1}(-u)=\tr K_+(u)K_-(u)\,.
\ll{tr}
\eq
Here an extra boundary $K$-matrix $K_+$ is some solution of a
"conjugated" reflection equation \cite{skl88,kusas93} on the
quantum space $\H_+$ defined in such way to guarantee the
commutativity $[\tau(u),\tau(v)]=0$. This transfer matrix gives
rise to the  hamiltonian and other integrals of
motion for a quantum system with the  space of states
$\H=\H_+\otimes\H_n\otimes\H_{n-1}\cdots\H_1\otimes\H_-$.

Looking for dynamical boundary matrices $K(u)$ we can start
with a single operator $L(u)$ in the chain (\ref{sklm}).  In
addition, we can begin either with the generalized reflection
equation (\ref{re2}) by using known twists $S(u,v)$
\cite{d90,ms96,kst96}, or with the exchange algebras
(\ref{res3}) by assuming some ansatz for the boundary matrix
$K(u)$.


\section{The Toda lattices}
\setcounter{equation}{0}
As an example, let us consider the
following $L$-operator
\bq
L(u)=\left(\begin{array}{cc}u-p&-\exp(q)\\
\exp(-q)&0\end{array}\right)\,,\qquad [p,q]=-i\eta\,,
\ll{ltoda}
\eq
where $(p,q)$ be a pair of canonical conjugated variables.
This $L$-operator is intertwined (\ref{fcr}) by the rational
Yang $R$-matrix
\bq R(u)=uI-i\eta P\,,
\qquad
P=\left(\begin{array}{cccc}1&0&0&0\\
0&0&1&0\\ 0&1&0&0\\0&0&0&1\end{array}\right)\,,\ll{rmat1}
\eq
where $P$ be permutation operator
in $\bC^2\otimes\bC^2$
and
\bq
R(u)=\left(\begin{array}{cccc}u-i\eta&0&0&0\\
0&u&-i\eta&0\\ 0&-i\eta&u&0\\0&0&0&u-i\eta\end{array}\right)=
\left(\begin{array}{cccc}f&0&0&0\\
0&g&h&0\\ 0&h&g&0\\0&0&0&f\end{array}\right)(u)\,,
\ll{rmat}
\eq
In that case the monodromy matrix $T(u)$ (\ref{tl}) describes
the Toda lattices associated with the root system $\A_n $
\cite{skl85a}.  The corresponding hamiltonian reads as
\bq
H_A=\sum_{j=1}^n \dfrac12 p_j^2 +\sum_{j=1}^{n-1}\exp(q_{j+1}-q_j)\,,
\ll{tan}
\eq
The set of the operators $\{L(q_j,p_j,u)\}_{j=1}^n$ (\ref{ltoda})
defines the monodromy matrix $T(u)$ (\ref{tl}), which is a
spin-1/2 representation of the Yangian $Y(sl(2))$ in $V=\bC^2$.
So, the inversion $T^i(u)=T^{-1}(-u)$ is equal to \cite{skl88}
\bq
T^i(u)=\s_2 T^t(-u-i\eta)\s_2/\Delta\{T(-u-\dfrac{i\eta}2)\}\,.
\ll{inv}
\eq
where $t$ means a matrix transposition, $\s_2$ is the Pauli
matrix and $\Delta\{T(u)\}$ is a quantum determinant of $T(u)$.
According to general recipe \cite{skl88}, it allows us to work 
with the algebra $\te{\U}_R$ instead of the $\U_R$.  The new 
algebra $\te{\U}_R$ has the following Sklyanin representation
\bq
K_-=T(u)K(u-\dfrac{i\eta}2)\s_2 T^t(-u)\s_2\,.\ll{ual}
\eq
Let us begin with the known scalar solution of the reflection
equation (\ref{re})
\bq
K_c(u)=\left(\begin{array}{cc}
{a}&{b}  \\
{c}&{d}\end{array}\right)(u)\equiv
\left(\begin{array}{cc}
{\a u+\d}&{\b u}  \\
{\g u}   &{-\a u + \d}\end{array}\right)\,,
\ll{kxc}
\eq
where $\a,\b,\g$ and $\d$ are complex numbers.
By using the monodromy matrix $T(u)$ (\ref{tl}), the usual
covariance property and solution $K_c(u)$ (\ref{kxc}) one can get the
monodromy matrix $K_-$ (\ref{ual}), which describes the Toda
lattices associated with the root system $\A_n $ by $\a=\d=\g=0$ (\ref{tan})
and with the root system $\B\C_n$ by $\b=1$
\cite{skl88,kuzts89a}.  At the second case
the corresponding hamiltonian is given by
\bq
H_{BC}=H_A-\dfrac{\g}2\exp(2q_1)+(\d-\a p_1)\exp(q_1)\,.
\ll{hbc}
\eq
Transfer matrix (\ref{tr}) with the "conjugated" to $K_c$
(\ref{kxc}) matrix $K_+$ allows to describe another Toda
lattices associated to the several affine root systems
\cite{bog76,skl88,kuzts89a}.

Looking for dynamical boundary matrix $K(u)$ let us begin with
the single operator $L(u)$ (\ref{ltoda}) in the chain and
introduce the following ansatz for the dynamical boundary
matrix
\bq
K(u)=\left(\begin{array}{cc}a&b\\
c&d\end{array}\right)(q,u)\,,
\ll{ak}
\eq
where $a,b,c$ and $d$ are functions of spectral parameter $u$
and one dynamical variable $q$ only.  This matrix $K(u)$
(\ref{ak}) depends on the half dynamical variables and,
therefore, may be constructed from the scalar solutions to
generalized reflection equation.

Inserting ansatz $K(q,u)$ (\ref{ak}) into the dynamical
exchange algebras (\ref{res3}) and then into the general
dynamical equation (\ref{res4}) we get two nontrivial upper
$c(q,u)=0$ and lower $b(q,u)=0$ triangular matrices.  In both
these solutions diagonal entries $a(u)$ and $d(u)$ are
independent on the dynamical variable $q$.

Note, triangular boundary
matrices with the property (\ref{trpr}) are obtained from the
dynamical equations (\ref {res4}-\ref{res3}) by using the
special ansatz (\ref{ak}). Only than, due to the special
structure of the boundary matrices, we can see that the
transformation (\ref{stw}) of the Yang solution $R(u)$
(\ref{rmat}) is just the Drinfeld twist (\ref{dtw}) depending
on the spectral parameters \cite{d90} only. Moreover, these
twists are closely connected to the twists of the underlying Lie
algebra $sl(2)$.


\subsection{\it Lower triangular dynamical matrix.}
\par\noindent
Inserting the lower triangular  matrix
\bq
K_d=\left(\begin{array}{cc}a(u)&0\\
c(q,u)&d(u)\end{array}\right)\,,\ll{kxd}
\eq
into the system (\ref{res3}) one gets two dynamical
equations
\bq [p,c(q,u)]=-z(q,u)d(u)\exp(q)\,,\qquad
   [p,c(q,v)]= z'(q,v)a(v)\exp(q)\,,\ll{ltdeq}
\eq
related with the following matrices
\ben
F=\left(\begin{array}{cccc}1&0&0&0\\
0&1&0&0\\ 0&0&1&0\\z(q,u)&0&0&1\end{array}\right)\,,
\quad G=\left(\begin{array}{cccc}1&0&0&0\\
0&1&0&0\\ 0&0&1&0\\z'(q,v)&0&0&1\end{array}\right)\,,\nn
\en
Here $z(q,u)$ and $z'(q,v)$ are functions of the spectral
parameters and of the dynamical variable $q$.
By using generators $\h,\e,\f$ of the underlying Lie algebra $sl(2)$
\bq [\h,\e]=2\e\,,\qquad [\h,\f]=-2\f\,,\qquad [\e,\f]=\h\,,\ll{slt}
\eq
let us introduce an appropriate element
$\F\in U(sl(2))\otimes U(sl(2))$
\[\F_z=exp(z\cdot\f\otimes \f)\,,\qquad z\in\bC\,.\]
belonging to a tensor product of the corresponding universal
enveloping algebras $U(sl(2))$ \cite{kst96}.  In the
fundamental spin-1/2 representation $\rho_{1\over2}$ we have
\[F =(\rho_{1\over2}\otimes \rho_{1\over2})\F_z\,,\qquad
G=(\rho_{1\over2}\otimes \rho_{1\over2})\F_{z'}\,.\]
However, now $z$ and $z'$ are the proper functions of both
spectral parameters. The associated twist (\ref{stw})
\ben
S(u,v)&=&FR(u+v)F^{-1}_{21}=I_{zz'}\Bigl[(u+v)I-i\eta P\Bigr]\,,\nn\\
I_{zz'}&=&exp\Bigl((z+z')\cdot\f\otimes \f\Bigr)\,,\nn
\en
is equal to
\[
S(u,v)=FR(u+v)G=
\left(\begin{array}{cccc}f&0&0&0\\
0&g&h&0\\ 0&h&g&0\\\Bigl[z(q,u)+z'(q,v)\Bigr]&0&0&f\end{array}\right)(u,v)\,.
\]
Next, the only possible solution to the dynamical equations
(\ref{ltdeq}) is given by
\[K_d=\left(\begin{array}{cc}u&0\\
u\g(q)&-u\end{array}\right)\,.\]
Here entry $\g(q)$ is defined by
\bq
[p,\g(q)]=z(q)\exp(q)\,,\ll{e1}
\eq
with an arbitrary function $z(q)$ of dynamical variable $q$ and
$z(q)=-z'(q)$, whence $S(u,v)=R(u+v)$.

Next, we have to solve the general dynamical equation
(\ref{res4}).  Let the matrix $S(u,v)$ is equal to
\[S(u,v)=
\left(\begin{array}{cccc}f&0&0&0\\
0&g&h&0\\ 0&h&g&0\\z(q,u,v)&0&0&f\end{array}\right)(u,v)\,,
\]
where $z(q;u,v)$ is an arbitrary entry.  Substituting this
matrix and the matrix $K_d$ (\ref{kxd}) into (\ref{res4})
one gets the following dynamical equation
\bq
a(v)[p,c(q,u)]+d(u)[p,c(q,u)]=
z(q;u,v)\dfrac{a(v)d(u)}{u+v-i\eta}\exp(q)\,,
\eq
which by $c(q,u)=\g(q)u$ reads as
\bq
[p,\g(q)]=z(q;u,v)\dfrac{a(v)d(u)}{(ua(v)+vd(u))
(u+v-i\eta)}\exp(q)\,.\ll{e2}
\eq
Solving the generalized reflection equation (\ref{re2}) with
this matrix $S(u,v)$ one gets the same possible nontrivial
solution $K_d$ (\ref{kxd}) only.

To construct the transfer matrix $\tau(u)$ (\ref{tr}) let us
substitute $q=q_1$ in the matrix $K_d(q,u)$ and  $q=-q_n$ in the
"conjugated" matrix $K_+(q,u)=K_d^t(-q_n,u)$.
Thus dynamical boundary matrices $K_d$ and $K_{d+}$ have
the common quantum spaces with the first $L_1$ and with the last
$L_{n}$ operators in the lattice. This generating polynomial
\ben
\tau(u)=\tr\Bigl( K_d^t(-q_n,u+{{i\eta}\over{2}})
L_n(q_n,u)\ldots L_1(q_1,u)\times\Bigr.\nn\\
\Bigl.\times K_d(q_1,u-{{i\eta}\over{2}})
\s_2 L_1^t(q_1,-u)\ldots L_n^{t}(q_n,-u)\s_2\Bigr)\nn
\en
has the form
\[\tau(u)=H_1u^{2n}+H_2u^{2n-2}+\ldots\]
and it gives rise the commutative family of $n$ functionally
independent integrals of motion.  If the entry
$z(u,v,q)$ is independent on dynamical variable $q$ then
solution to equations (\ref{e1}) and (\ref{e2})  is equal to
\[\g(q)=\g\exp(q)+\b\,,\]
whereas first integral $H_1$ has the following factorable form
\[
H_1=J_1\cdot J_n=\left(2p_1+\g_-e^{2q_1}+\b_-e^{q1}\right)e^{q_1}\cdot
e^{-q_n}\left(-2p_n+\g_+e^{-2q_n}+\b_+e^{-qn}\right)\,.
\]
Here $(\g_-,\,\b_-)$ and $(\g_+,\,\b_+)$ are the
free parameters associated to the boundary
matrices $K_d(q_1,u)$  and  $K_d^t(-q_n,u)$, respectively.

This integrable system may be considered as
the constrained hamiltonian system either  with one
constraint $H_1=const$,
or  with two constraints
\[q_1=const_1\,,~q_n=const_n\,,
\qquad {\rm or}\qquad J_1=const_1\,,~J_n=const_n\,.\]


\subsection{\it Upper triangular dynamical matrix.}
\par\noindent
Inserting upper triangular  matrix
\bq
K_D=\left(\begin{array}{cc}a(u)&b(q,u)\\
0&d(u)\end{array}\right)\,,\ll{kxdd}
\eq
into the system (\ref{res3}) one gets two dynamical
equations
\bq [p,b(q,u)]=-w(q,u)\,a(u)\exp(q)\,,\qquad
   [p,b(q,v)]= -w'(q,v)\,d(v)\exp(q)\,,\ll{deq}
\eq
related with the following matrices
\bq
F=\left(\begin{array}{cccc}1&0&0&0\\
0&1&w(q,u)&0\\ 0&0&1&0\\0&0&0&1\end{array}\right)\,,\qquad
G=\left(\begin{array}{cccc}1&0&0&0\\
0&1&0&0\\ 0&w'(q,v)&1&0\\0&0&0&1\end{array}\right)\,.\ll{utfg}
\eq
Here $w(q,u,v)$ and $w'(q,u,v)$ are functions of spectral
parameters and dynamical variable $q$.

In generators (\ref{slt}) the corresponding twist element
$\F\in U(sl(2))\otimes U(sl(2))$ is equal to
\[\F_w=exp(w\cdot\f\otimes\e)\,,\qquad w\in\bC\,\]
(see factorization of the universal $R$-matrix
in \cite{kst96, ms96}).
In the fundamental spin-1/2 representation $\rho_{1\over2}$ we have
\[F =(\rho_{1\over2}\otimes \rho_{1\over2})\F_w\,,\qquad
G=P(\rho_{1\over2}\otimes \rho_{1\over2})\F_{w'}P\,.\]
However, now $w$ and $w'$ are the corresponding to (\ref{deq})
functions of the spectral parameters. We can see,
$G=F_{21}^{-1}$ up to change the twist parameters $w(q,u)\to
-w'(q,v)$.  The associated twisted matrix $S(u,v)$ (\ref{stw})
is equal to
\[
S(u,v)=
F_{12}(w)\Bigl((u+v)I-i\eta P\Bigr) F_{21}^{-1}(w')
=J_{ww'}-i\eta I_{ww'}P\,,\]
where
\ben
J_{ww'}&=&(\rho_{1\over2}\otimes \rho_{1\over2})
\exp(w\cdot\f\otimes\e+w'\cdot\e\otimes\f)\,,\nn\\
I_{ww'}&=&(\rho_{1\over2}\otimes \rho_{1\over2})
\exp\Bigl((w+w')\cdot\f\otimes\e\Bigr)\,.\nn
\en
In fact, matrix $S(u,v)$ has the following form
\bq
S(u,v)=\left(\begin{array}{cccc}f&0&0&0\\
0&g+h(w+w')+ww'g&h+wg&0\\ 0&h+w'g&g&0\\0&0&0&f\end{array}\right)(u,v)\,,
\ll{smat}
\eq
where $f,\,g$ and $h$ are entries of the initial $R$-matrix
(\ref{rmat}).

For a given $S$-matrix (\ref{smat}) we shall not solve
dynamical equations (\ref{deq}) and  generalized reflection
equation (\ref{re2}) in generic. Here we restrict ourselves to
those particular solutions which are related to interesting
physical systems only.

At first, let us introduce  upper triangular matrix $K_D$
with the following entries
\bq
a(u)=d(u)=u\,,\quad b(q,u)=\b\exp(q)+\g\,,\ll{1s}
\eq
One immediately gets
\[w(u)=i\eta\b u^{-1}\,,\qquad w'(v)=i\eta\b v^{-1}\,,\]
where parameters $w$ and $w'$ of the twist are independent on
the dynamical variable $q$. Thus, from the linear
matrix-function $R(u)$ (\ref{rmat}) we obtain the rational
matrix $S(u,v)$, which has the upper and lower triangular
residues at the points $u=0$ and $v=0$, respectively.

The second more complicated solution $K_{gD}$ is defined by
\bq
a(u)=u^2+\a u+\d\,,\quad d(u)=a(-u)\,,\quad
\quad b(q,u)=(\b\exp(q)+\g)u\,.\ll{2s}
\eq
For the both solutions
\[K(u)K(-u)=\mp\phi(u)I\,,\]
where function $\phi(u)$ is equal to the determinant of
the matrices $K_D$ and $K_{gD}$, respectively.

To construct the transfer matrix $\tau(u)$ (\ref{tr}) let us
substitute $q=q_1$ in the matrix $K_D(q,u)$ and  $q=-q_n$ in the
"conjugated" matrix $K_+(q,u)=K_D^t(-q_n,u)$.
It means that dynamical boundary matrices $K_D$ and $K_{D+}$
have the common quantum spaces with the first $L_1$ and the
last $L_n$ operators in the lattice.  The generating polynomial
$\tau(u)$ (\ref{tr})
\ben
\tau(u)=\tr\Bigl( K_D^t(-q_n,u+\dfrac{i\eta}2)\,
L_n(q_n,u)\cdots L_1(q_1,u)\Bigr.\times\nn\\
\Bigl.\times K_D(q_1,u-\dfrac{i\eta}2)\,
\s_2 L_1^t(q_1,-u)\cdots L_n^{t}(q_n,-u)\s_2\Bigr)\nn
\en
has the form
\bq
\tau(u)=H_1u^{2n+2}+H_2u^{2n}+\ldots\,,\ll{dexp}
\eq
and it gives rise to the commutative family of $n+1$
functionally independent integrals of motion. The use of  the
second dynamical matrix $K_{gD}$ (\ref{2s}) leads to the
similar results.

For the both solutions first integral $H_1$ into expansion
(\ref{dexp}) may be considered as the factorable constraint
\[
H_1=J_1\cdot J_n=\left(-2e^{q_1}+\b_-e^{q_1}+\g_-\right)\cdot
\left(2e^{-q_n}+\b_+e^{-qn}+\g_+\right)\,.
\]
Here $(\g_-,\,\b_-)$ and $(\g_+,\,\b_+)$ are the free
parameters associated to dynamical entries of the boundary
matrices $K_D(q_1,u)$  and $K_{D+}=K_D^t(-q_n,u)$,
respectively.

In contrast with the lower triangular solution this constraint
is easy removed, if the parameters $w$ and $w'$ of the twist
are independent on the dynamical variable $q$.  Namely, unless
otherwise indicated, set
\[\b_\pm=\mp 2\,,\]
such that $H_1=\g_-\g_+=const$ and
generating polynomial (\ref{dexp}) gives rise to $n$
independent integrals of motion only.

After canonical transformation of the variables
\bq e^q\to 1-ch(q)\,,\ll{ctr}
\eq
in the first $\H_1$ and in the last $\H_n$ local quantum spaces
in the chain, the associated hamiltonians $H$ read as
\ben
H_D&=&H_A+\exp(-q_{n-1}-q_n)+\exp(q_1+q_2)\,,\nn\\
\ll{dnt}\\
H_{gD}&=&H_D
+\dfrac{\a_1}{\sinh^2{{q_1}\over2}}
+\dfrac{\a_2}{\sinh^2 q_1}
+\dfrac{\a_3}{\sinh^2{{q_n}\over2}}
+\dfrac{\a_4}{\sinh^2 q_n }\,.\nn
\en
Here four constants $\a_j$ are functions of the four initial
constants $\a_\pm$ and $\d_\pm$ in diagonal
entries (\ref{2s}) of the boundary matrices \cite{kuzts89a}.

Thus, dynamical boundary matrix $K_D$ (\ref{1s}) corresponds to
the Toda lattices associated with the root system $\D_n $
\cite{bog76}. Since, the single spin-1/2 representation $T(u)$
(\ref{tl},\ref{ltoda})  of the Yangian $Y(sl(2))$ may be used to
construct the monodromy matrices for the Toda lattices
associated to all the classical infinite series of the root
systems. The second solution $K_{gD}$ (\ref{2s}) allows to
add four extra parameters in the hamiltonian $H_D$ and it is
known Inozemtsev's generalization of the Toda system
\cite{in89}.

Boundary matrices
$K'_D=L(u)K_DL^{-1}(-u)$ and $K'_{gD}=L(u)K_{gD}L^{-1}(-u)$
were at first found in \cite{kuzts89a} starting from the known
$2n\times 2n$ Lax matrices \cite{in89}.  They are solutions to
the usual reflection equation, which have the ultralocal
commutation relations with other matrices in the chain.
Note, we have to use the two different representations
$T(u)$ (\ref{tl}) of the Yangian $Y(sl(2))$ to describe
Toda lattices associated with the $\B\C_n$ and $\D_n$ root systems by
using matrices $K_c$ and $K'_{D}$, respectively.  In the
classical mechanics factorization $K'_D=L(u)K_D(q,u)L^{-1}(-u)$
on the terms with non-\-ul\-tra\-lo\-cal commutator relations
has been applied to separation of variables in \cite{kuz97}.

Two outer automorphisms of the space of infinite-dimensional
representations of the Lie algebra $sl(2)$ be used to recover
these boundary matrices $K'_D$ and $K'_{gD}$ in \cite{ts96b}.
It would be interesting to study interrelations among these
automorphisms of $sl(2)$ and the twists (\ref{dtw}) of the
usual rational $R$-matrix.

The relativistic Toda lattices associated to the $\D_n$ root
systems \cite{kuzts93a} may by easy embedded in the proposed
scheme as well. In that case the corresponding $R$-matrix is
the known trigonometric solution to the Yang-Baxter equation
\cite{kuzts93a} and the associated twist is connected to the
twist of the algebra $sl_q(2)$.


\subsection{\it The Drinfeld twist and separation of variables method.}
\par\noindent
We know that the representation theory of the Drinfeld twists
may be very useful in the framework of the algebraic Bethe
ansatz \cite{ms96}.  According to \cite{ms96}, for the $XXX$-$1
\over 2$ and $XXZ$-$1 \over 2$ Heisenberg (inhomogeneous)
quantum spin chains of finite length $n$ associated
$\F$-matrices diagonalize the generating matrix of scalar
products of quantum states of these models. They also
diagonalize the diagonal (operator) entries of the quantum
monodromy matrix.

Now we shortly discuss interrelations of the Drinfeld twists
and the separation of variables method proposed by Sklyanin
\cite{skl95}.  For sake of brevity we shall work with the
corresponding classical objects.

Starting from the known $2n\times 2n$ Lax matrices we can get
solutions $K'_D$ or $K'_{gD}$ to the reflection equation
(\ref{re}) with ultralocal commutation relations. In the
classical mechanics the associated $2\times 2$ Lax matrix
is equal to
\ben
\L'(u)&=&{K'_D}^t(-q_n,u)\,
L_{n-1}(q_{n-1},u)\cdots L_2(q_2,u)\times\nn\\
&\times&K'_{D}(q_1,u)\,
\s_2 L_2^t(q_2,-u)\cdots L_{n-1}^{t}(q_{n-1},-u)\s_2\,.
\ll{lax'}
\en
Here all the local matrices $L_j(u),~j=2,\ldots,n-1$ and boundary matrices
$K'_{D}(q_{1,n},u)$ are defined on the different phase spaces such that
$\{\on{L_j}{1}(u),\on{K'_{D}}{2}(q_{1,n},v)\}=0$.

In the Sklyanin approach \cite{skl95} the use of this Lax
matrix forces to apply the dynamical normalization of the
corresponding Baker-Akhiezer vector-function $\Psi$
\cite{kuz97}. Recall, the choice of the proper normalization
$\vec{\a}$ \cite{skl95}
\[(\vec{a},\Psi)=\sum_{j=1} \a_j(u)\Psi_j(u)=1\]
allows us to fix special analytical properties of this
meromorphic eigenfunction $\Psi$ of the Lax matrix $\L'(u)$
\[\L'(u)\Psi=z\Psi\,.\]
An appropriate normed vector-function $\Psi$ has to
possess the necessary number of poles in involution and all the
extra poles of $\Psi$ be constants \cite{skl95}.  However, one
does not usually know the separating normalization in advance.

It is clear, using a similarity transformation for the Lax
matrix $\L'(u)$
\bq
\L'(u)\to \L(u)=V(u)\L'(u)V^{-1}(u)\,,\ll{simtr}\eq
the any normalization $\vec{a}$ may be turns into the simplest
constant normalization vector \cite{skl95}.  So, in the
classical mechanics the problem is to find an appropriate
similarity transformation (\ref{simtr}).

In the quantum mechanics an action of the similarity
transformations (\ref{simtr}) may be transferred into the
$R$-matrix level. If we permit to transform $R$-matrix in the
special twists only, we essentially restrict the
freedom related to similarity transformations (\ref{simtr}).
The twisted $R$-matrix enjoys most the properties of the
usual $R$-matrix \cite{d90} and possible that it gains some new
properties.  As an example, admissible \cite{d90} or
factorizing \cite{ms96} twists may be used.

In this paper, starting from the known Yang solution $R(u)$ of
the Yang-Baxter equation we introduce twist $S(u,v)$, which has
another analytical properties. Next, solving the corresponding
dynamical equations and the generalized reflection equation we
get dynamical boundary matrices $K_D$ or $K_{gD}$.
The associated $2\times 2$  Lax matrix is equal to
\ben
\L(u)&=&K_{D}^t(-q_n,u)\,
L_{n}(q_{n},u)\cdots L_1(q_1,u)\times\nn\\
&\times&K_{D}(q_1,u)\,
\s_2 L_1^t(q_1,-u)\cdots L_{n}^{t}(q_{n},-u)\s_2\,.
\ll{lax}
\en
Here two local matrices $L_{1,n}(u)$ and boundary matrices
$K_D(q_{1,n},u)$ are defined on the common phase spaces and the
corresponding Poisson brackets relations
$\{\on{L_{1,n}}{1}(u),\on{K_D}{2}(q_{1,n}v)\}\neq 0$ may be derived
from the quantum algebras (\ref{res3}).

Two Lax matrices $\L'(u)$ (\ref{lax'}) and $\L(u)$ (\ref{lax})
are related by the similarity transformation (\ref{simtr}).
Now $V(u)$ is equal to the inversion matrix $V(u)=L^i_n(q_n,-u)$
(\ref{inv}).  This transformation leads to the appearance of
the twist $S(u,v)$ (\ref{smat}) instead of the Yang
solution $R(u+v)$ in the quantum algebraic relations.

Although matrices $K_D$ and $K_{gD}$ are solutions
to the more complicated generalized reflection equation (\ref{re2})
and dynamical equations (\ref{res4}), nevertheless these matrices make
possible to use the simplest constant normalization of the
associated Baker-Akhiezer function \cite{skl95} in the
classical mechanics. Namely, for the Toda lattice associated to
the $\D_n$ root system in the classical mechanics the Lax
matrix $\L(u)$ (\ref{lax}) has the following matrix form
\bq
\L(u)=\left(\begin{array}{cc}A&B\\
C&D\end{array}\right)(u)\,.\ll{lax1}
\eq
Choosing the simplest constant normalization $\vec{\a}=(1,0)$
of the associated Baker-Akhiezer function \cite{skl95} the
separation variables $\{x_j\}_{j=1}^n$ are defined as zeroes of
the entry $B(u)$ \cite{kuz97}
\[B(u=x_j)=0\,,\]
according to general recipe \cite{skl95}.  It is easy to prove
that they are real eigenvalues of the symmetric matrix defined
by recursion, which has been proposed in \cite{kuzts89a}.
Other separation variables are sitting on the spectral curve of
the Lax matrix (\ref{lax1}) with the defined above variables
$\{x_j\}_{j=1}^n$ \cite{skl95,kuz97}.

Thus, in the considered above example, the use of the Drinfeld
twists in the quantum case leads to the suitable Lax
representation and to the simplest separating normalization in
the classical mechanics. We can see that the algebraic properties of
the Lax matrix relate to the analytical properties
of its eigenfunction.  Of course, we have to study the
suitable properties of this twist, which correspond to the
separation of variables.  Note, the constant normalization of
the Baker-Akhiezer function allows to develop the quantum
counterpart of the classical separation of variables method
within the $R$-matrix approach \cite{skl85a} for the Toda
lattices.


\section{Integrable tops closed to the Toda lattice.}
\setcounter{equation}{0}

\par\noindent
Let variables $l_i, g_i, ~i=1, 2, 3$ be generators
 of the Lie algebra $e(3)$ with commutator relations
 \ben
 &&\bigl[ l_i\,, l_j\,\bigr]= -i\eta\e_{ijk}\,l_k\,,
 \qquad \bigl[ l_i\,, g_j\,\bigr]= -i\eta\e_{ijk}\,g_k\,,\nn\\
 &&\bigl[ g_i\,, g_j\,\bigr]= 0\,, \qquad i, j=1, 2, 3,
 \nn
 \en
and with the following Casimir operators
\[J_1=(g, g)\,,\qquad J_2=(l, g)\,.\]
Let us introduce quantum operator $T(u)$ for the Neumann's
system
\bq
 T\,(u)=
 \left(\begin{array}{cc}
 u^2-2l_3\,u - l_1^2-l_2^2-\dfrac14~~~&
 i(g_+\,u-\dfrac12\{g_3\,, l_+\,\})\\
 i(g_-\,u-\dfrac12\{g_3\,, l_-\,\})&
 g_3^2\end{array}\right)\,,
 \ll{qvop}
\eq
here braces $\{, \}$ mean an anticommutator.  Operator $T(u)$
(\ref{qvop}) at the level $J_2=(l, g)=0$ obeys the fundamental
commutator relations (\ref{fcr}) with the rational $R$-matrix
(\ref{rmat}) and closely related to the Toda system \cite{kuzts89}.

The use of the usual covariance property (\ref{t-}), operator
$T(u)$ (\ref{qvop}) and the constant boundary matrices $K_c$
(\ref{kxc}) allows to describe the quantum
Ko\-wa\-lew\-ski-Chap\-ly\-gin-Go\-rya\-chev top
\cite{kuzts89}.

Let us consider the known constant solution (\ref{1s}) or
(\ref{2s}) to the generalized reflection equation (\ref{re2})
with the considered above twisted matrix $S(u,v)$ (\ref{smat}).
By using this solution we may introduce another solution of the
same equation (\ref{re2}), which is the function defined on the
Abel subalgebra of the $e(3)$.  Thus, we can get dynamical
boundary matrices on the common quantum space with operator
$T(u)$ (\ref{qvop})
\ben
&&K_-=\left(\begin{array}{cc}u^2+\a_-u+\d_-&iu(\b_- g_++\g_-)\\
0&u^2-\a_-u+\d_-\end{array}\right)\,,\nn\\
\ll{ktop}\\
&&K_+=\left(\begin{array}{cc}u^2+\a_+u+\d_+&0\\
iu(\b_+ g_-+\g_+)&u^2-\a_+u+\d_+\end{array}\right)\,,\nn
\en
such that
\[\on{K_+}1\on{K_-}2=\on{K_-}2\on{K_+}1\,.\]
If we set \[\b_\pm=-2\,,\]
then the generating polynomial (\ref{tr}) is equal to
\[\tau(u)=\tr\Bigl[ K_+T(u)K_-\s_2 T^{t}(-u)\s_2\Bigl]=
u^6 (2J_1-\g_+\g_-) + u^4 J_3 + u^2 J_4\,.\]
Two independent integrals of motion $J_3$ and $J_4$
are mutually commute at the level $J_2=(l, g)=0$. Here
we present the corresponding hamiltonian $J_3$  in the classical
mechanics only
\ben
J_3&=&l_+ l_- (g_+ \g_+ -\g_+ \g_- +g_- \g_-)
-2 l_3^2(2 g_--\g_+) (2 g_+-\g_-)\nn\\
\nn\\
&+&2 l_3\Bigl[
2J_1(\a_++\a_-)-(\a_- \g_+ g_++\a_+ \g_- g_-)\Bigr]\nn\\
\nn\\
&+&l_3g_3 \Bigl[2l_- (2 g_+-\g_-)+2l_+ (2 g_--\g_+)
-J_3(\a_++\a_-)
\Bigr]\nn\\
\nn\\
&+&g_3 (\g_+ \a_- l_++\a_+ \g_- l_-)
+\d_+ \g_- g_- + \d_- \g_+g_+\,.\nn
\en
By analogy with (\ref{ctr}), the use of automorphisms
of the Lie algebra $e(3)$ might allow to rewrite this hamiltonian
in the more physical form.  For us  it is more
important that all three matrices in the chain are defined
on the single common quantum space.

\section{Conclusions}
\setcounter{equation}{0}
We discuss dynamical boundary matrices defined on
the common quantum space with other operators in the chain.
These matrices are solutions to generalized reflection equation
and dynamical exchange equations. These equations include
usual $R$-matrix and its Drinfeld twist depending on the spectral
parameters. As an example, we consider the twist of the Lie
algebra $sl(2)$  related to the Toda lattices associated to the
$\D_n$ root system.

This work was partially supported by RFBR grant.

\end{document}